\documentclass[aps,prd,twocolumn,showpacs,preprintnumbers,amsmath,amssymb,footinbib]{revtex4-1}
\usepackage[dvips]{color,graphicx}
\usepackage{dcolumn}
\usepackage{bm}
\usepackage{slashed}
\usepackage{amssymb,amsmath}
\usepackage{mathrsfs}
\usepackage{overpic}
\usepackage[varg]{txfonts}
\newcommand{\ba}{\begin{array}}
\newcommand{\ea}{\end{array}}

\newcommand{\ie}{{i.e.}}

\begin{document}
\title{Chiral crystallization in an external magnetic background\\
-- Chiral spiral versus Real kink crystal --\\
}
\author{Hiroaki Abuki}%
\email[E-mail:~]{abuki@auecc.aichi-edu.ac.jp}
\affiliation{%
Department of Education, Aichi University of Education,
1 Hirosawa, Kariya 448-8542, Japan}
\date{\today}
\begin{abstract}
We study how an external magnetic field modifies the chiral phase
 structure of QCD, in particular the phases characterized by
 inhomogeneous chiral condensates.
The magnetic field can be systematically incorporated into a
 generalized Ginzburg-Landau framework, and it turns out to induce
 a model independent universal coupling between the magnetic field
 and the axial isospin current.
The resulting effect is found to be drastic especially in the chiral limit; 
no matter how small the magnetic intensity is, 
 the tricritical Lifshitz point is totally washed out, and
the real kink crystal is replaced by a \emph{magnetically induced} chiral spiral.
The current quark mass, on the other hand, has an opposite effect,
protecting the chiral critical point from the magnetically induced
 chiral spiral.
But once the magnetic intensity exceeds a critical value, the critical
 point no longer exists.
We draw a semiquantitative conclusion that the critical point
 disappears for $\sqrt{eB}\agt 50$~MeV.

\end{abstract}
\pacs{12.38.Mh, 21.65.Qr}
\maketitle
\section{Introduction}
There has recently been a growing interests in possible crystal structures
formed in dense QCD matter governed by the strong interaction
\cite{Nakano:2004cd,Nickel:2009ke,Buballa:2014tba}.
The effect of magnetic field on QCD matter has also been the subject of
intensive studies.
Phenomenologically, exploring possible forms of strongly interacting
matter under the magnetic field is relevant to the physics of
magnetars; the compact stellar objects known to have a strong magnetic
field $B\sim 10^{11}$~T at its surface, and an even stronger field
$B\sim 10^{14}$~T might be realized in their cores
\cite{Shapiro1991,Duncan:1992hi,Ferrer:2010wz}.

How the magnetic field affects the chiral symmetry breaking
in homogeneous QCD matter has been a matter of active debates
for last decades.
Phenomenological models predict the magnetic catalysis
\cite{Suganuma:1990nn,Klevansky:1989vi,Klimenko:1991he,Gusynin:1994re}
while lattice QCD simulations show the opposite;
the inverse magnetic catalysis \cite{Bali:2012zg,Endrodi:2015oba}.
The mechanism for the magnetic catalysis is rather transparent, but that
for the inverse one still seems to lack a common consensus
\cite{Bruckmann:2013oba,Fukushima:2012kc,Chao:2013qpa,Preis:2010cq,Andersen:2014xxa}.

The effect of an external magnetic field is even more significant for
inhomogeneous matter \cite{Frolov:2010wn,Tatsumi:2014wka,Cao:2016fby,%
Nishiyama:2015fba,Andersen:2017lre,Andersen:2018osr}.
For example, in \cite{Tatsumi:2014wka} the authors have shown that
the magnetic field strongly stabilizes the chiral spiral ($\chi$-spiral)
 aka the dual chiral density wave (DCDW) in the chiral limit, and as a
consequence it brings about a new critical point on the temperature axis
in the phase diagram.
On the other hand, the effect of current quark mass is known to favor
the real kink crystal phase (RKC), but the chiral spiral might survive
as the ``massive dual chiral density wave'' where the complex phase of
condensate is skewed from a linear function of space coordinate
\cite{Yoshiike:2015wud}.

In this article, we study the effect of magnetic fields on the chiral
phases with a particular focus put on how it modifies the
(inhomogeneous) phase structure in the vicinity of the critical point.
Several studies are already devoted on how the magnetic field
affects the critical point itself
\cite{Ruggieri:2014bqa,Costa:2015bza,Rechenberger:2016gim}.
We concentrate on the phase structure in the neighborhood of the critical
point.
We first show that it is possible to derive systematically the
generalized Ginzburg-Landau (gGL) action in the presence of an external
magnetic field ($\bm{B}$) without specifying any details of spatial form
of the chiral condensate.
We find a $\bm{B}$-odd term which couples to the axial isospin current.
This term is considered to be universal in the sense that it is
independent of any model parameters such as coupling strength and cutoff.
We evaluate the strength of the coupling as a function of $\mu$ and
$T$ for an arbitrary intensity of magnetic field.
We then study the impact of $\bm{B}$-odd term on the phases near the
critical point. 
It turns out that the term strongly favors the chiral
spiral, the condensate accompanied by a complex phase, bringing a
dramatic change in the phase diagram.

\vspace*{1ex}
\noindent
\section{Deriving generalized Ginzburg-Landau action.}
\begin{figure}[t]
\begin{center}
\includegraphics[width=65mm]{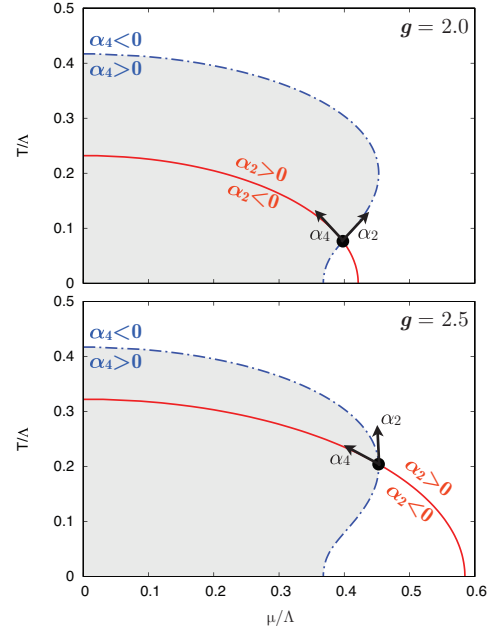}
\caption{The illustrative figure that shows how $(\alpha_2,\alpha_4)$
 spans the local coordinate in the vicinity of the TCP within the NJL
 type model.
 This is depicted for two values of coupling $g\equiv
 G\Lambda^2=2.0$ (upper panel) and $g=2.5$ (lower panel).
The solid line (red online) shows the curve on which $\alpha_2=0$, while
the dotted-dashed line (blue online) does that for $\alpha_4=0$.
The point of intersection gives the location of the TCP.
The region for $\alpha_4>0$ is shaded. 
The solid line in the shaded area represents the second-order chiral
 transition, while that in the unshaded area only gives the spinodal
 line staying still in the broken phase, and once $\alpha_2<0$
 the symmetric phase with $\sigma=0$ starts to constitute a local minimum.
 The figure was taken from Ref.~\cite{Abuki:2013vwa}.
}
\label{fig:map}
\end{center}
\end{figure}

In any quark-based model, the generalized Ginzburg-Landau (gGL) action
density in the absence of the external magnetic field can be derived in
the same way as described in \cite{Abuki:2013pla,Abuki:2013vwa,Abuki:2011pf}.
For two-flavor, the quark loop contribution to the effective action can
be expanded in the powers of the quark self-energy
$\Sigma({\bf x})=m_q+\sigma({\bf x})+i\gamma_5\bm{\tau}\cdot\bm{\pi}({\bf x})$
as, up to the second order in $\Sigma$,
\begin{equation}
 \delta S_\mathrm{eff}=\frac{T}{2}\sum_n\sum_{{\bf x},\,{\bf y}}\mathrm{tr}\left[%
S^{(0)}(i\omega_n,{\bf x}-{\bf y})\Sigma({\bf y})S^{(0)}(i\omega_n,{\bf y}-{\bf
x})\Sigma({\bf x})\right].
\label{eq:expansion}
\end{equation}
Here $S^{(0)}(i\omega_n,{\bf x})=-\int{d{\bf p}}e^{i\bf{p}\cdot{\bf x}}%
\frac{i\omega_n\gamma_0-{\bf p}\cdot{\bm{\gamma}}}{\omega_n^2+{\bf
p}^2}$ is the quark propagator with $\omega_n=\pi T(2n-1)$ being the
Matsubara frequency. 
Expressing $\Sigma({\bf y})=\Sigma({\bf
x})+\sum_{i=1}^\infty\frac{1}{i!}\left[%
({\bf y}-{\bf x})\cdot\bm{\nabla}\Sigma({\bf x})\right]^i$, we can perform
a systematic derivative expansion of the effective action.
Writing the action with the gGL action density $\omega$ as 
$S_{\mathrm{eff}}=\int d{\bf
x}\omega({\bf x})$, the result is found up to the sixth order in
$\sigma$, $\pi_a$ ($a=1,2,3$) and $\bm{\nabla}\equiv\bm{\partial}_{\bf
x}$ as
\begin{equation}
\begin{array}{c}
 \omega({\bf x})=%
\displaystyle -h\sigma+\frac{\alpha_2}{2}\phi^2%
+\frac{\alpha_4}{4}\left(\phi^4+(\bm{\nabla}\phi)^2\right)%
+\frac{\alpha_6}{6}\bigg(\phi^6+\frac{1}{2}(\Delta\phi)^2\\[2ex]%
\qquad\qquad\;\;+3[\phi^2(\bm{\nabla}\phi)^2-(\phi\cdot\bm{\nabla}\phi)^2]%
+5(\phi\cdot\bm{\nabla}\phi)^2\bigg),
\end{array}
\label{eq:gGL}
\end{equation}
where we have switched to the chiral four-vector notation
$\phi=(\sigma,\bm{\pi})$.
$h$ and $\alpha_{n}$ ($n=2, 4, 6$), are the gGL couplings which depend on
quark chemical potential $\mu$ and temperature $T$.
The first term $-h\sigma$, which we call ``$h$-term'' hereafter, is
responsible for the explicit symmetry breaking by the current quark mass
$m_q$.
In fact, $h$ is actually proportional to $m_q$:~
\begin{equation}
h=m_q(8N_c)T\sum_n\int\frac{d\bf p}{(2\pi)^3}%
 \frac{1}{(\omega_n-i\mu)^2+{\bf p}^2},
\end{equation}
where $N_c=3$ specifies a number of colors.
The integral is divergent in ultraviolet and needs some regularization
to be evaluated.
In the spirit of the gGL approach, we simply take $h$ as a parameter
characterizing the explicit symmetry breaking.
Similarly the expressions for $\alpha_n$ can be found as
\begin{equation}
\alpha_{2i}=\frac{\delta_{i,1}}{2G}+8N_cT\sum_n\int\frac{d{\bf
 p}}{(2\pi)^3}\frac{(-1)^i}{((\omega_n-i\mu)^2+{\bf p}^2)^i},
 \label{eq:alphas}
\end{equation}
for $i=1, 2, \cdots$.
It is only $\alpha_2$ that has an extra tree-level counter contribution
$1/2G$ with $G$ being a four-Fermi coupling in the
standard NJL model \cite{Nickel:2009ke}.
$\alpha_2$ and $\alpha_4$ should be vanishing at the tricritical point (TCP)
$(\mu_{\mathrm{TCP}} ,T_\mathrm{TCP})$ which is expected to show up in
the phase diagram in the chiral limit $h=0$.
In order to illustrate how $(\alpha_2,\alpha_4)$ maps onto the $(\mu,T)$
phase diagram in the chiral limit within the NJL-type model, we show
in Fig.~\ref{fig:map} the lines for $\alpha_2=0$ and $\alpha_4=0$ for two
different values of coupling $G$.
The intersection of these two curves indicated by a circle represents the
location of TCP.
From the figure, we can see how the GL couplings $(\alpha_2,\alpha_4)$
spans a local coordinate in the vicinity of TCP.

We now move onto the effect of external magnetic field.
There is a direct effect on the quark propagator.
It is easy to expand quark propagator in the powers of magnetic field
along with the line described in \cite{Gorbar:2013uga}. 
At the leading order in $\bm{B}$,
\begin{equation}
 S(i\omega_n,{\bf p})=S^{(0)}(i\omega_n,{\bf p})%
+(QB_i)\frac{{\slashed p}_\parallel+{\slashed \mu}}{[(i\omega_n+\mu)^2%
-{\bf p}^2]^2}\frac{i\epsilon_{ijk}\gamma^j\gamma^k}{2},
\label{eq:propinB}
\end{equation}
where we have used the four vector notation $p^\mu_\parallel%
=(i\omega_n+\mu,{\bf p}_{\parallel})$ with ${\bf
p}_\parallel=({\bf p}\cdot\bm{B})\bm{B}/|\bm{B}|^2$ being
the momentum component parallel to the magnetic field.
$Q=\mathrm{diag.}(2e/3,-e/3)$ is the electric charge matrix in the
flavor space.
Plugging Eq.~(\ref{eq:propinB}) into $S^{(0)}$ in the integrand of
Eq.~(\ref{eq:expansion}),
and extracting the term linear 
in $\bm{B}$ in the gGL action density, we have a
universal (anomalous) coupling \cite{Abuki:2016zpv}:
\begin{equation}
 \delta\omega_B({\bf x})=\frac{1}{4N_f}\frac{\partial\alpha_4}{\partial\mu}
e\bm{B}\cdot\left(%
\sigma\bm{\nabla}\pi_3-\pi_3\bm{\nabla}\sigma\right).
\label{eq:anom}
\end{equation}
$\mu$-derivative of the fourth gGL coefficient $\alpha_4$ can be evaluated
without any UV divergence.
$$
 \frac{1}{4N_f}\frac{\partial\alpha_4}{\partial\mu}%
=-\frac{N_c}{8\pi^2 T}f(e^{-\mu/T}),
$$
where 
$
f(e^{-y})=\frac{1}{2\pi}\mathrm{Im}\psi^{(1)}\left(\frac{1}{2}%
-i\frac{y}{2\pi}\right),
$
with $\psi^{(1)}$ the trigamma function.
Several remarks are in order here.
(i)~The same result was obtained in quite a different manner in
\cite{Tatsumi:2014wka} where only the lowest Landau level (LLL) was
taken into account. 
In our approach the full set of Landau levels are incorporated,
and this procedure gives the identical result.
This is because the spectral asymmetry lies only in the LLL.
(ii)~This term was derived for the specific ansatz, \ie, DCDW in
\cite{Tatsumi:2014wka}.
In our approach, the spatial profile of condensate in three
space dimensions is not postulated to be in any specific form.
(iii)~A similar term is obtained also at zero temperature 
\cite{Son:2007ny}.

\begin{figure}[tb]
\centering
\begin{overpic}[width=7cm,clip]{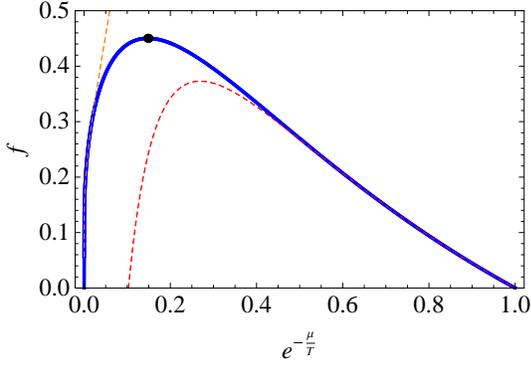}
\end{overpic}
\caption{$f$ as a function of the fugacity $z=e^{-\mu/T}$. $z=1$
 corresponds to $\mu=0$, while $z=0$ represents the high density limit
$\mu\to\infty$.
The circle placed on the curve is the point where $f$ takes maximum.
}
\label{fig:0}
\end{figure}

We display in Fig.~\ref{fig:0}, $f$ as a function of fugacity
$z=e^{-\mu/T}$ by a solid curve.
Dashed curves correspond to the following approximations:
\begin{equation}
f(e^{-\mu/T})=\begin{cases}%
\frac{7\zeta(3)}{2\pi^2}%
\frac{\mu}{T}-\frac{31\zeta(5)}{4\pi^4}%
\left(\frac{\mu}{T}\right)^2+\cdots &\mbox{for $\mu\ll T$,}\\[2ex]
\frac{T}{\mu}+\frac{\pi^2}{3}\left(\frac{T}{\mu}\right)^2+\cdots&%
\mbox{for $\mu\gg T$.}
\end{cases}
\label{eq:app}
\end{equation}
$f$ takes maximum value $0.45$ at the point $\mu/T\cong 1.91$ which
is marked by a circle placed on the solid curve in
Fig.~\ref{fig:0}.

In principle, in order to evaluate the magnitude of universal coupling,
we need to fix $\mu/T$ at the exact location of TCP because
we are working in the neighborhood of TCP within gGL framework.
However, this does depend on the detail of model, so we define here
\begin{equation}
\bm{b}\equiv \frac{N_c}{8\pi^2 T}f(e^{-\mu/T})(e\bm{B}),
\end{equation}
call 
$\delta\omega_B=-\bm{b}\cdot(\sigma\bm{\nabla}\pi_3-\pi_3\bm{\nabla}%
\sigma)$ ``$\bm{b}$-term''.
We regard $\bm{b}$ as as a gGL coupling constant whose magnitude
serves as a measure of the intensity of the external magnetic field.
From the symmetry viewpoint, $h$-term only breaks the chiral symmetry
to the isospin $\mathrm{SU}(2)$,
while the $\bm{b}$-term explicitly breaks several symmetries: the time
reversal symmetry, the rotational symmetry, in addition to
the isospin $\mathrm{SU}(2)$ symmetry which is to be broken down to
$\mathrm{U}_\mathrm{Q}(1)$.

Switching to the complex notation for the condensate
$\Delta=\sigma+i\pi_3$, the gGL action density can be cast into the
form
\begin{equation}
\begin{array}{rcl}
\omega({\bf
 x})&=&
-\bm{b}\cdot\mathrm{Im}\left[\Delta^*\bm{\nabla}\Delta\right]%
-h\mathrm{Re}\left[\Delta\right]\\[2ex]
&&+\frac{\alpha_2}{2}|\Delta|^2+\frac{\alpha_4}{4}%
\left(|\Delta|^4%
+|\bm{\nabla}\Delta|^2\right)+\frac{\alpha_6}{6}\Big(%
 |\Delta|^6\\[2ex]
&&+3|\Delta|^2|\bm{\nabla}\Delta|^2
+2\left(\mathrm{Re}[\Delta^*\bm{\nabla}\Delta]\right)^2%
+\frac{1}{2}|\bm{\nabla}^2\Delta|^2%
\Big).
\end{array}
\label{eq:delta}
\end{equation}
The first two terms are the symmetry breaking sources, 
responsible for the current quark mass and the magnetic field,
respectively.
It can be easily guessed that the $h$-term favors the real condensate,
while the $\bm{b}$-term stabilizes the complex condensate such as the
chiral spiral.
We note that our $\bm{b}$-term is exactly in the same form as the one
obtained in one-dimensional (1D) Gross-Neveau model \cite{Boehmer:2007ea},
where it was shown that the spiral phase dominates the phase diagram.
This term is forbidden in the three dimensional NJL model because it
breaks the rotational symmetry.
The magnetic field induces this term even in 3D so that it opens the
 possibility that the complex condensate comes into play in the QCD
 phase diagram.

\begin{figure}[tb]
\centering
\begin{overpic}[width=7.9cm,clip]{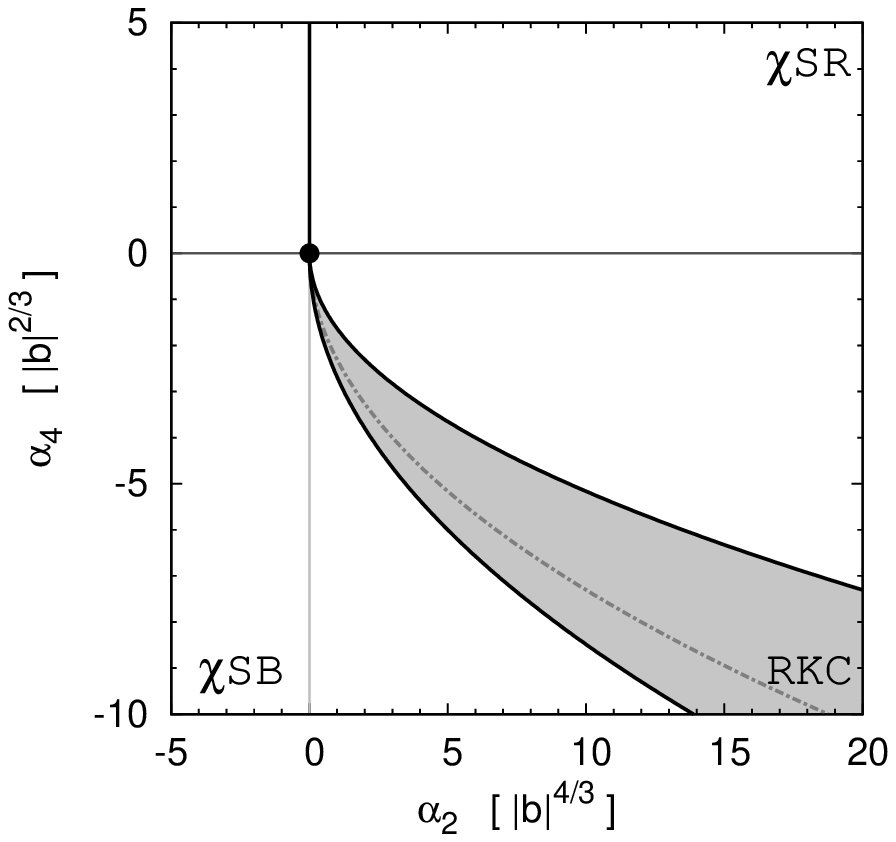}
\end{overpic}
\begin{overpic}[width=7.9cm,clip]{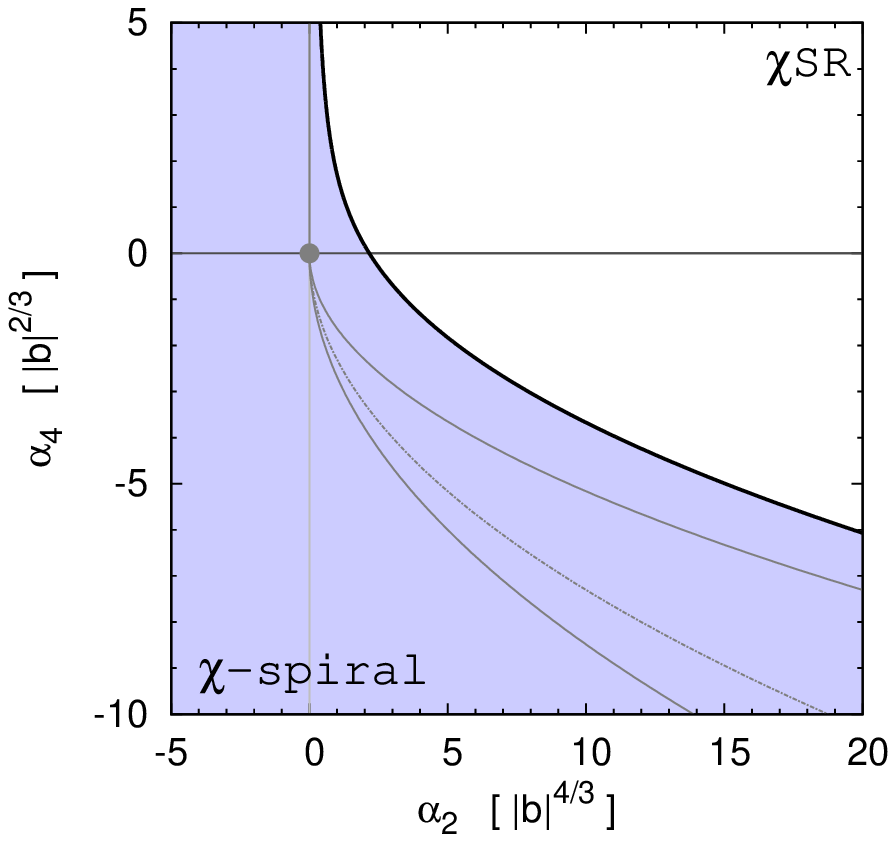}
\end{overpic}
\caption{The phase diagrams in the chiral limit, $h=0$. 
Upper panel:~The phase diagram for zero magnetic field. 
Lower panel:~The phase diagram for nonvanishing magnetic field $\bm{b}$.}
\label{fig-1}
\end{figure}

 Apart from the above new coupling appearing at the leading (linear)
 order in $e\bm{B}$, there should also be corrections to the original
 GL couplings $\alpha_2$, $\alpha_4$, etc.
 However, the corrections to these terms should start from the quadratic
 order in $e\bm{B}$ because of the rotational invariance.
 In the Appendix~\ref{eq:exdemo}, we demonstrate that this is actually the
 case by performing the explicit gradient expansion to $\alpha_2$ and $\alpha_4$.
 In principle these corrections bring about a shift of the location of TCP.
 However, this effect occurs at order $(eB)^2$, and in the present study
 we focus on the effects at the leading order in $(eB)$ on to the phase
 diagram.

 \vspace*{1ex}
 \noindent
\section{How do magnetic fields modify the phase diagram?}
Let us first begin with the chiral limit.
This corresponds to ignoring the $h$-term in the gGL energy
density (\ref{eq:gGL}).
We measure every dimensionful quantity with the proper power of
$(\alpha_6)^{-1/2}$.
Then, we can also scale out the effect of $\bm{b}$, by taking
$|b|^{4/3}$ ($|b|^{2/3}$) for the unit of $\alpha_2$ ($\alpha_4$).
The phase diagram for $|b|=0$ is depicted in the upper panel
of Fig.~\ref{fig-1}.
First, note that the Lifshitz tricritical point (TCP)
is located at the origin which, in principle, has a unique map onto the
$(\mu_{\mathrm{TCP}},T_\mathrm{TCP})$ in QCD phase diagram.
Second, the real kink crystal (RKC) condensate characterized by the spatial
form \cite{Abuki:2013pla,Abuki:2013vwa,Nickel:2009wj}:
$$
\textstyle
\sigma({\bf x})%
=k\mathrm{sn}(a,\nu)\left(\nu^2\mathrm{sn}({\bf k}\cdot{\bf x},\nu)\mathrm{sn}({\bf k}\cdot{\bf x}+a,\nu)%
+\frac{\mathrm{cn}(a,\nu)\mathrm{dn}(a,\nu)}{\mathrm{sn}^2(a,\nu)}\right),
$$
enters in between the chiral symmetric phase ($\chi$SR) and the chiral symmetry
broken phase ($\chi$SB).
One might wonder why $|b|$ comes in the units of
$\alpha_2$ and $\alpha_4$ in spite of zero magnetic field $b=0$.
This is just for a convenience, and in this case $|b|$ is arbitrary. 
In fact, the phase boundaries are independent of $|b|$, 
since any critical lines are expressed by $\alpha_4^2\propto\alpha_2$.
\begin{figure}[t]
\centering
\begin{overpic}[width=5.5cm,clip]{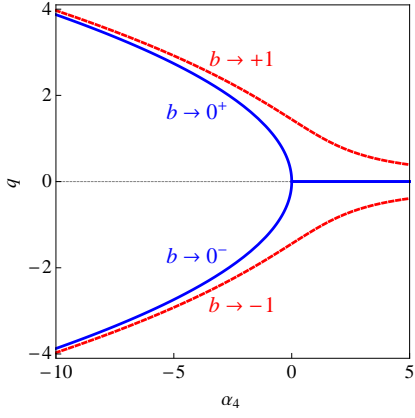}
\end{overpic}
\caption{The wavenumber $q$ in the unit of $1/\sqrt{\alpha_6}$
at the onset of instability to the formation of chiral condensate
as a function of $\alpha_4$.
The solid (blue online) lines represent the $b=0$ case, while the 
dashed (red online) lines represent the $b=\ne0$ case.
The formation of chiral density wave is signaled by $q\ne 0$ branch.
}
\label{fig:wavenumber}
\end{figure}

In the lower panel of Fig.~\ref{fig-1}, the phase diagram for nonvanishing
$|b|$ is displayed.
The phase structure is completely changed by the emergence
of a complex chiral spiral, $\Delta({\bf x})=\Delta_0
e^{i{\bm{q}}\cdot{\bf x}}$, 
denoted by ``$\chi$-spiral'' in the figure.
In this phase the direction of $\bm{q}$ is locked to the direction of
the magnetic field.
TCP is killed by the stabilization of the $\chi$-spiral phase, and
there is only a second order phase transition line between $\chi$SR
and $\chi$-spiral phases.
We stress that this drastic change happens for an arbitrary intensity of
magnetic field.
It means that the standard $\chi$SB phase becomes unstable against the
formation of density wave, and TCP will never be realized in the
presence of magnetic field.

We here briefly try to figure out why the TCP is totally washed out
by the magnetic field with an arbitrary strength.
In order to make this point clear, we here make the stability analysis
of the symmetric phase with respect to the formation of chiral condensate.
Since we are interested in the border line separating the symmetric phase and
the broken phase, we take the ansatz
$\Delta=\sigma+i\pi_3=\Delta_0e^{i\bm{q}\cdot{\bf x}}$;
this covers both the homogeneous condensate ($q=0$) and the chiral spiral
($q\ne 0$) which is known to be degenerate with the RKC on the phase
boundary.
Then making every dimensionful quantity dimensionless by $1/\sqrt{\alpha_6}$,
the thermodynamic potential becomes
$$
\Omega=\left(\frac{\alpha_2}{2}+\frac{\alpha_4q^2}{4}+\frac{q^4}{12}%
-\bm{b}\cdot\bm{q}\right)\Delta_0^2+\left(%
\frac{\alpha_4}{4}+\frac{5q^2}{12}\right)\Delta_0^4+\frac{1}{6}\Delta_0^6.
$$
The effect of the magnetic field appears through $(-\bm{b}\cdot\bm{q}%
\Delta_0^2)$ term, which is linear in ${\bm{q}}$.
The favorable configuration is the alignment of $\bm{q}$ into the
direction of magnetic field, $\bm{q}=q\hat{b}$.
Solving the stationary condition $\frac{\partial\Omega}{\partial q}=0$
results in $q$ as some algebraic function of $\alpha_4$, $\Delta_0^2$, and $b$.
There are two extreme cases where the naive expansion about $b$ becomes
possible; (i) $\alpha_4\ll 0$ and (ii) $\alpha_4\gg 0$.
Let us start with the case (i) where $\alpha_4\ll 0$.
In this case, we have
$$
 q=\left(\sqrt{\frac{3|\alpha_4|}{2}}+\frac{b}{|\alpha_4|}\right)%
+\frac{5\left(4b-\sqrt{6|\alpha_4|^3}\right)}{12\alpha_4^2}\Delta_0^2+\cdots.
$$
Plugging this into $\Omega$ results in
$$
\Omega=\left(\frac{\alpha_2}{2}-\frac{3\alpha_4^2}{16}-\sqrt{\frac{3|\alpha_4|}{2}}b\right)\Delta_0^2%
+\left(\frac{3}{8}|\alpha_4|+\frac{5b}{2\sqrt{6|\alpha_4|}}\right)\Delta_0^4+\cdots.
$$
Then we see that the second order phase transition takes place at
\begin{equation}
 \frac{\alpha_2}{\alpha_4^2}%
=\frac{3}{8}+\frac{\sqrt{6}b}{|\alpha_4|^{3/2}}
  +{\mathcal O}\left(\big(b|\alpha_4|^{-3/2}\big)^2\right),
  \label{eq:cline1}
\end{equation}
and the wavenumber that the instability develops is
\begin{equation}
q=\sqrt{\frac{3|\alpha_4|}{2}}\left(1+\sqrt{\frac{2}{3}}\frac{b}{|\alpha_4|^{3/2}}\right).
 \label{eq:k1}
\end{equation}
We see that the critical wavenumber increases because of the magnetic field.
Equation (\ref{eq:cline1}) explains the magnetic shift of critical line from
the unperturbed one [$\alpha_2=\frac{3}{8}\alpha_4^2$ $(\alpha_4<0)$]
to the positive $\alpha_2$ direction, which is actually what is seen in
Fig.~\ref{fig-1}.
In the case (ii) where $\alpha_4\gg 0$, solving the stationary condition
for $q$ yields
$$
q=\frac{2b}{\alpha_4}-\frac{10b}{\alpha_4^2}\Delta_0^2+\cdots.
$$
Substituting the above expression into $\Omega$, and performing
expansion about $\Delta_0$, we arrive at
$$
\Omega=\left(\frac{\alpha_2}{2}-\frac{b^2}{\alpha_4}\right)\Delta_0^2%
+\left(\frac{\alpha_4}{4}+\frac{5b^2}{3\alpha_4^2}\right)\Delta_0^4%
\cdots.
$$
From these we see that the system has an instability to the formation of
density wave at
\begin{equation}
 \frac{\alpha_2}{\alpha_4^2}=\frac{b^2}{\alpha_4^3}%
  +{\mathcal O}\left(\big(b\alpha_4^{-3/2}\big)^3\right),
\end{equation}
where the wavenumber where the instability sets in is
\begin{equation}
 q=\frac{2b}{\alpha_4}.
  \label{eq:k2}
\end{equation}
It is remarkable that $q\ne0$ for $b\ne 0$ so that the transition is always
from the symmetric phase to density wave with magnetically induced
small wavenumber. 
This is quite different from $b=0$ case where
the system has an instability to homogeneous condensate, as is seen in
the upper panel of Fig.~\ref{fig-1}.

From the discussion above, we notice that the expansion in $b$ breaks
down when $|\alpha_4|$ becomes small because the combination
$b/|\alpha_4|^{3/2}$ becomes large.
Then we need to solve the instability condition numerically.
Figure \ref{fig:wavenumber} shows the numerically obtained critical
wavenumber as a function of $\alpha_4$.
We see that the both situations ($\alpha_4\ll0$ and
$\alpha_4\gg0$) are well explained by the analytic formulas,
Eqs.~(\ref{eq:k1}) and (\ref{eq:k2}).
For $b=0$, we see that $\alpha_4=0$ (TCP) is realized as a bifurcation point
across which the inhomogeneous phase with $q\ne 0$ develops.
Sign of $q$ there may be determined by the direction of an infinitesimal
magnetic field.
The point $\alpha_4=0$ is actually the second order phase transition
from homogeneous to inhomogeneous condensate.
For $b\ne 0$, we notice that the instability always occurs at a
nonvanishing momentum $q\ne0$. 
This means that the transition is always from the symmetric phase to the
density wave irrespective of the value of $\alpha_4$.
This is in fact the case as we see in the lower panel of Fig.~\ref{fig-1}.
There is no longer TCP in the phase diagram for $b\ne 0$, because the $b$-term acts as
an external field to make the condensate inhomogeneous in space, and this external
field smears the bifurcation structure (equivalently the second
order phase transition) at $\alpha_4=0$ as seen in Fig.~\ref{fig:wavenumber}.
This is why an infinitesimal magnetic field washes the TCP totally out from the phase
diagram.

\begin{figure}[tb]
\centering
\begin{overpic}[width=7cm,clip]{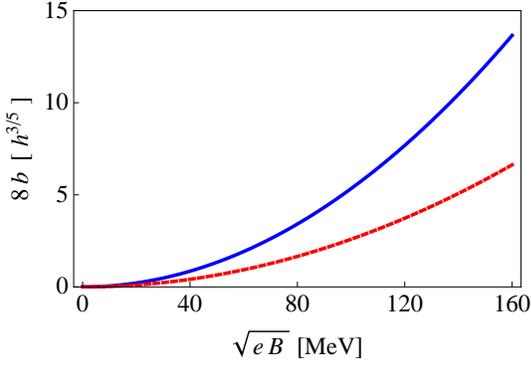}
\end{overpic}
\caption{$8b/h^{3/5}$ as a function of $eB$ in the physical unit MeV.
The solid curve corresponds to $T=50$ MeV, while the dashed one stands
 for $T=100$ MeV.
Each curve should be considered to be the upper limit for corresponding
temperature because it is evaluated at the fugacity parameter at which
 $b$ takes maximum.
}
\label{fig:2}
\end{figure}

\vspace*{1ex}
\noindent
\section{Quark mass versus magnetic field}
Next we consider the effect of current quark mass $h$ together
with the magnetic field $\bm{b}$.
In this case, making every dimensionful quantities dimensionless
by taking $\sqrt{1/\alpha_6}$ as a natural energy unit,
and then performing a scale transformation $\Delta\to h^{1/5}{\Delta}$,
$\nabla_{\bf x}\to h^{1/5}\nabla_{\bf x}$ (that is ${\bf x}
\to h^{-1/5}{\bf x}$), the (dimensionless)
gGL energy density Eq.~(\ref{eq:delta}) becomes
\begin{equation}
 \begin{array}{rcl}
h^{-{6/5}}\omega({\bf
 x})&=&
-h^{-3/5}\bm{b}\cdot\mathrm{Im}\left[\Delta^*\bm{\nabla}\Delta\right]%
-\mathrm{Re}\left[\Delta\right]\\[2ex]
&&+\frac{h^{-4/5}\alpha_2}{2}|\Delta|^2+\frac{h^{-2/5}\alpha_4}{4}%
\left(|\Delta|^4%
+|\bm{\nabla}\Delta|^2\right)+\frac{1}{6}\Big(%
 |\Delta|^6\\[2ex]
&&+3|\Delta|^2|\bm{\nabla}\Delta|^2
+2\left(\mathrm{Re}[\Delta^*\bm{\nabla}\Delta]\right)^2%
+\frac{1}{2}|\bm{\nabla}^2\Delta|^2%
\Big).
 \end{array}
\end{equation}
This means that the phase diagram only depends on three parameters
$\alpha_2/h^{4/5}$, $\alpha_4/h^{2/5}$ and $b/h^{3/5}$, and these are
nothing but $\alpha_2$, $\alpha_4$ and $b$ measured in proper units,
$h^{4/5}$, $h^{2/5}$, and $h^{3/5}$, respectively.
So when $h\ne 0$, the important combinatorial parameter is $b/h^{3/5}$
which determines how large the magnetic effect on to the phase structure is.
When this parameter is large, the effect of magnetic field prevails over
that of current quark mass.
This parameter is a linear function of the magnetic intensity $eB$
and is inversely proportional to $m_q^{3/5}$, \ie, $b/h^{3/5}%
\propto eB/m_q^{3/5}$.
The value of $b/h^{3/5}$ should have one to one correspondence to
$eB$ once $m_q$, $\mu_{\mathrm{TCP}}$, and $T_{\mathrm{TCP}}$ are all set.
Just for a guide, we show in Fig.~\ref{fig:2} the parameter $8\times
(b/h^{3/5})$
as a function of $eB$, for $T=50$~MeV (a solid curve) and for $T=100$~MeV (a
dashed curve).
For evaluation, we took a choice $\Lambda=600$~MeV for a momentum cutoff,
$m_q=5$~MeV and fixed $\mu=1.91\,T$ corresponding to the point indicated
by circle in Fig.~\ref{fig:0}.
Therefore the corresponding curves should be considered to be the upper
limits.

\begin{figure*}[tb]
\centering
\begin{overpic}[width=7.9cm,clip]{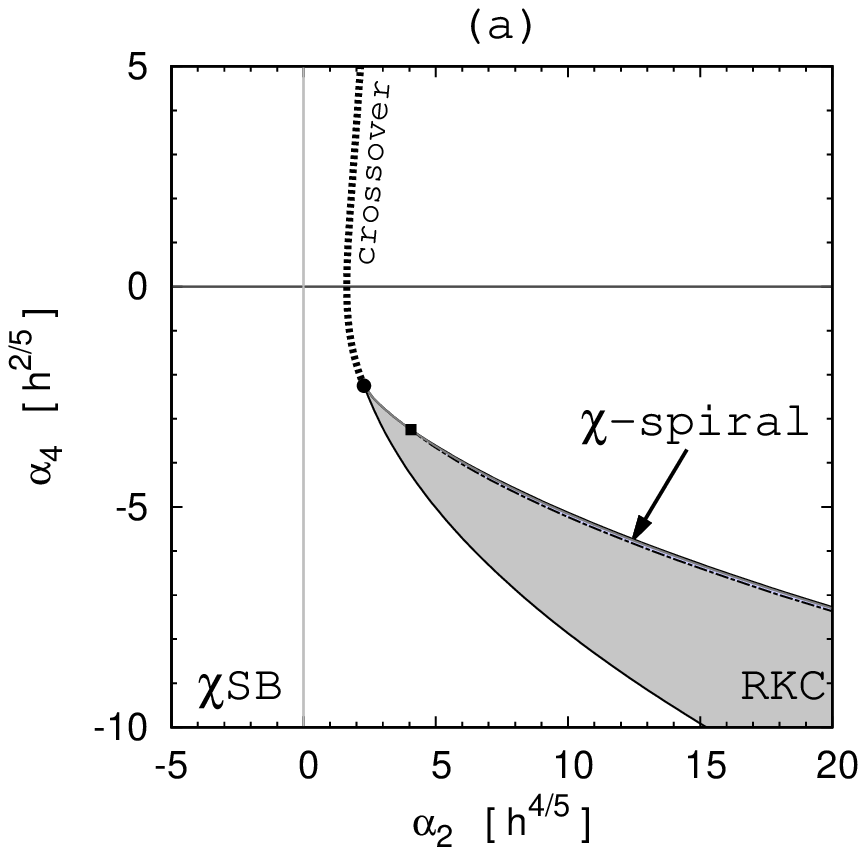}
\end{overpic}
\begin{overpic}[width=7.9cm,clip]{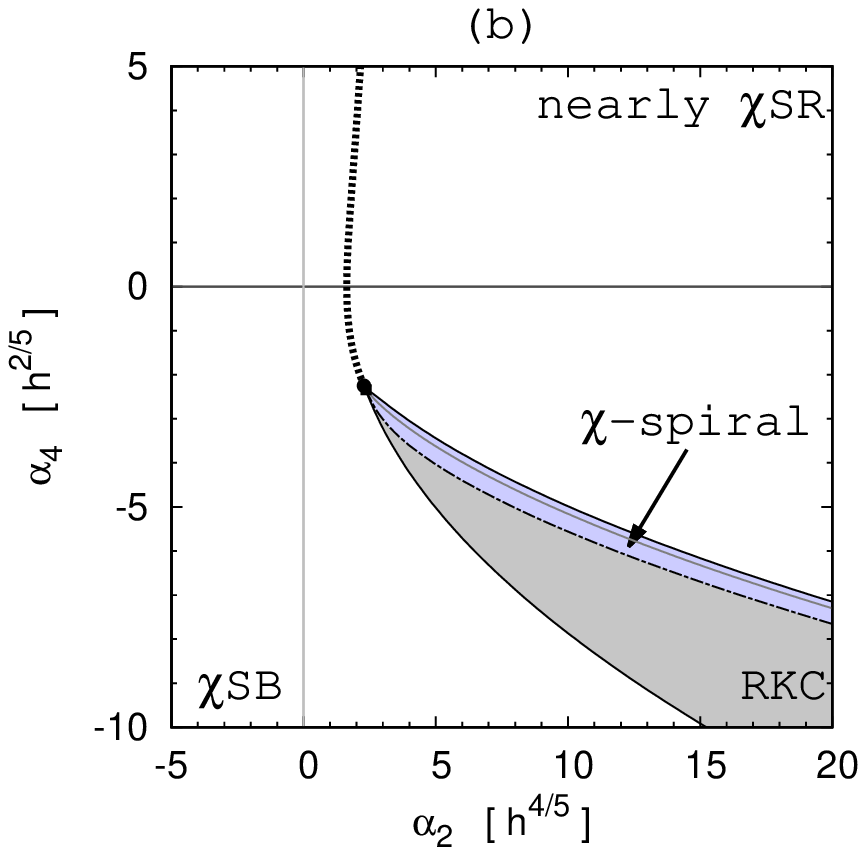}
\end{overpic}

\vspace*{3ex}

\begin{overpic}[width=7.9cm,clip]{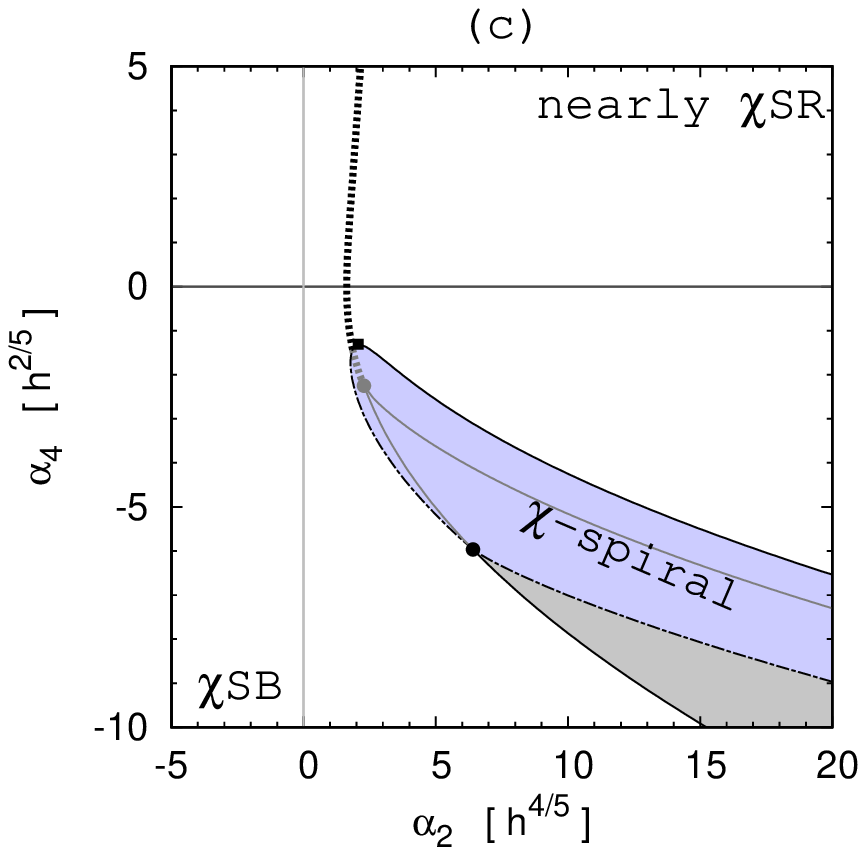}
\end{overpic}
\begin{overpic}[width=7.9cm,clip]{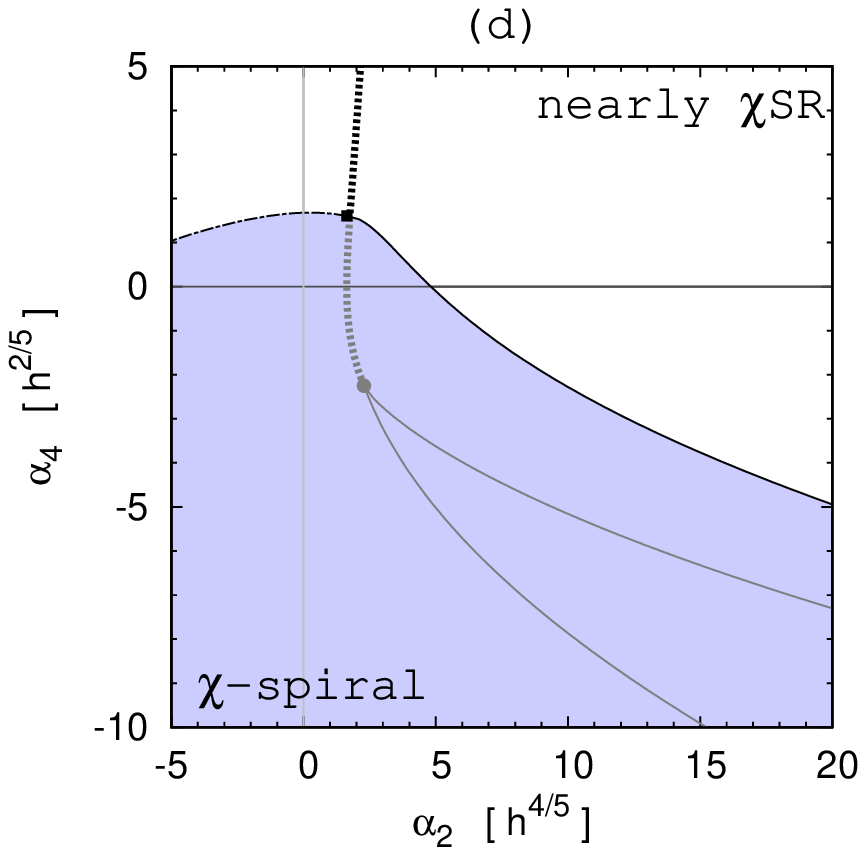}
\end{overpic}
\caption{The phase diagrams off the chiral limit.
(a):~$8b=0.2\times h^{3/5}$ [$\sqrt{eB}\sim 20-30$~MeV]. (b):~$8b=1.0\times
 h^{3/5}$ [$\sqrt{eB}\sim 40-60$~MeV].
(c):~$8b=5.0\times h^{3/5}$ [$\sqrt{eB}\sim
 100-130$~MeV]. (d):~$8b=15\times h^{3/5}$ [$\sqrt{eB}\sim 170-230$~MeV].
}
\label{fig-2}
\end{figure*}

In Fig.~\ref{fig-2} we show the phase diagrams for four different
values of magnetic field.
Displayed in Fig.~\ref{fig-2}(a) is for $8b=0.2\times h^{3/5}$,
where the effect of $\bm{b}$ is relatively weaker than that of the
current quark mass ($h$-term).
Note, however, even in this case the magnetic energy is quite large in
physical unit; the estimated value is $\sqrt{eB}\sim 20-30$~MeV
corresponding to $B\sim 7\times10^{12}-10^{13}$~T.
We see that the phase diagram is not much modified at this magnetic
intensity.
The magnetic field replaces only a tiny thin region near the phase
boundary between the $\chi$SR and RKC phases with a \emph{modified}
$\chi$-spiral defined by $\Delta=M_0+\Delta_0e^{i\bm{q}\cdot{\bf x}}$
with $M_0$, $\bm{q}$ and $\Delta_0$ the variational parameters.
This is magnetically induced chiral spiral.
However, a major part of the RKC and the Lifshitz critical point
itself remain intact.
The Lifshitz critical point is hereinafter referred to as
``critical point'' or ``CP'' simply.
We conclude that the current quark mass plays a role to protect
the critical point and the RKC phase from a weak magnetic field.
Figure~\ref{fig-2}(b) presents the phase diagram for $8b=1.0\times
h^{3/5}$, that roughly corresponds to $\sqrt{eB}\sim 40-60$~MeV ($B\sim
(3-6)\times 10^{13}$~T).
At this magnetic intensity, we see a sizable region for the modified
$\chi$-spiral.
Accordingly CP is killed and replaced by a new critical point where
the second order phase transition from the $\chi$-spiral
to the $\chi$SR turns into a first order one from the $\chi$-spiral
to the RKC (or $\chi$SB).
In Fig.~\ref{fig-2}(c), the phase diagram for a stronger magnetic field
$8b=5.0\times h^{3/5}$ is depicted. 
This corresponds to $\sqrt{eB}\sim 100-130$~MeV
($B\sim (1-3)\times 10^{14}$~T).
The region for the modified $\chi$-spiral gets significantly magnified,
and it now completely covers the original critical point.
There is a new critical point, denoted by a black square, where the
second order phase transition at which the $\chi$-spiral ends at
large $\alpha_2$ side, changes into a first order one at small
$\alpha_2$ side.
Figure~\ref{fig-2}(d) represents the phase diagram at an even stronger
magnetic field $8b=15\times h^{3/5}$, that is estimated to be
$\sqrt{eB}\sim 170-230$~MeV ($B\sim (5-9)\times 10^{14}$~T).
In this extreme case, the effect of magnetic field completely 
dominates over that from $h$-term.
The RKC phase is replaced by the $\chi$-spiral, which now spreads
over a wide region.
We see that the new critical point still exists on the phase boundary,
where the second order phase transition turns into a first order one.

\vspace*{1ex}
\noindent
\section{Conclusion and outlook}
We studied the effects of an external magnetic field on the chiral phase
structure of QCD within the generalized Ginzburg-Landau (gGL) effective
action.
We first derived the gGL action performing a derivative expansion up to
the sixth order in condensates and spatial derivatives.
Expanding the action also up to the lowest nontrivial order in a current
quark mass and a magnetic field, we obtained the explicit symmetry
breaking sources, $h$-term and $\bm{b}$-term, respectively.
The $h$-term explicitly breaks the chiral symmetry to the diagonal
isospin $\mathrm{SU(2)}$, while the $\bm{b}$-term violates the time
reversal symmetry, and reduces the isospin $\mathrm{SU(2)}$ down to
$\mathrm{U}_{\mathrm{Q}}(1)$, the spatial rotational symmetry
$\mathrm{SO(3)}$ down to $\mathrm{O(2)}$, the rotation about the
magnetic axis.
It is clearly seen in the obtained gGL action that these two
symmetry breaking terms have competing effects on the condensate; the
former prefers the real condensate, while the latter favors the complex
condensate spatially modulated in the direction of magnetic field.
We have computed the phase diagrams for nonvanishing magnetic fields.
In the chiral limit, the effect of an external magnetic field is such
drastic that it completely washes out the tricritical point as well as
the real kink crystal (RKC) phase.
There is only a second order phase transition at which the chiral spiral
phase ($\chi$-spiral) terminates.
On the other hand, the effect of current quark mass was found to
protect the RKC phase and the chiral critical point from being
invaded by a magnetically induced $\chi$-spiral.
However, as the intensity of magnetic field increases, the $\chi$-spiral
phase gradually eats the coast region of the high density boundary
between the RKC and nearly symmetric phases.
When the magnetic field strength exceeds a critical value, 
the effect of magnetic field prevails over that of current quark mass,
and the critical point and RKC phase get completely
replaced by the $\chi$-spiral phase.
We estimated the critical magnetic field ranging in between
$\sqrt{eB}\sim 40-60$~MeV depending on the location of the critical point.
We confirmed that, in the regime of strong magnetic fields,
the shape of the phase structure approaches the extreme one obtained in
the chiral limit.

There are several possible directions of extension of current work.
First, it is interesting to go beyond the standard gGL expansion,
for example along the line described in \cite{Carignano:2017meb}.
Second, nonequilibrium dynamics, in particular the relaxation dynamics
of inhomogeneous condensates under the magnetic field is worth to be
explored \cite{Carlomagno:2018ogx}.
This is because the $b$-term explicitly breaks the time reversal symmetry
so is expected to bring a sizable effects on the time dependence of the
condensates.
Lastly, it would be important and interesting subject to see if the chiral
spiral is also stabilized by a magnetic field in nuclear matter
\cite{Heinz:2013hza,Takeda:2018ldi}.

\vspace*{1ex}
\noindent
{\it Acknowledgements. --}
The author thanks R. Yoshiike and T. Tatsumi for useful discussions.
This work was supported by JSPS KAKENHI Grant Number JP16K05346.
The numerical calculations were carried out on Cray XC40 at 
the Yukawa Institute for Theoretical Physics, Kyoto University.

\appendix
\vspace*{1ex}
\section{Quark propagator in a magnetic background}
In this section, we briefly review how to derive the expression
Eq.~(\ref{eq:propinB}) following the procedure described in
\cite{Gusynin:1994re}.
The fermion propagator in the presence of a magnetic field can
be separated into two parts, the Schwinger phase
breaking the translational invariance, and the translationally
invariant part:
\begin{equation}
S(x,y)=e^{i\Phi(x,y)}\bar{S}(x-y).
\end{equation}
Let the direction of $\bm{B}$ be the $z$-direction ($\bm{B}=(0,0,B)$)
and $Q$ be the electric charge of the fermion, the Schwinger phase
in the Landau gauge $A_\mu(x)=(0,0,Bx_1,0)$ can be computed as
\begin{equation}
\Phi(x,y)=-Q\int_y^xdz_\mu A^\mu(z)=\frac{QB}{2}(x_1+y_1)(x_2-y_2),
\end{equation}
where the integration path is just a straight line connecting 
$y$ and $x$.
For the remaining part $\bar{S}(x-y)$, introducing the momentum $p$
and expanding in the Landau levels, we arrive at the expression
\begin{equation}
 \bar{S}(i\omega_n,p_\parallel,\bm{p}_\perp)=2^{-\ell^2\bm{p}_\perp^2}%
  \sum_{k=0}^\infty\frac{(-1)^kD_k(p)}{(i\omega_n+\mu)^2-2k/\ell^2-p_\parallel^2-m^2},
\end{equation}
where $\ell=1/\sqrt{|QB|}$ is the magnetic length, $m$ is the fermion mass,
$\mu$ is the chemical potential, $\omega_n=\pi T(2n+1)$ with $n$ being
an integer is the Matsubara frequency, $p_\parallel=p_z$ and
$\bm{p}_\perp=(p_x,p_y)$ is the momentum component perpendicular to the
magnetic field.
$P_\pm$ is the spin projection onto the direction of magnetic field,
defined by
$$
P_\pm=\frac{1}{2}\left(\bm{1}\pm\mathrm{sgn}(QB)i\gamma^1\gamma^2\right).
$$
Introducing the four-momentum notation $p^\mu=(i\omega_n,\bm{p}_\perp,p_\parallel)$,
the expression of $D_k(p)$ is obtained as follows.
\begin{equation}
 \begin{array}{rcl}
 D_k(p)&=&\bigg\{\left(\slashed{p}_\parallel+\slashed{\mu}+m\right)%
	       \left[P_+L_k^0(2\ell^2\bm{p}_\perp^2)-P_-L_{k-1}^0%
		(2\ell^2\bm{p}_\perp^2)\right]\\[2ex]
	       & &+(2\bm{p}_\perp\cdot\bm{\gamma}_\perp)%
	       L^1_{k-1}(2\ell^2\bm{p}_\perp^2)\bigg\},
 \end{array}
\end{equation}
where $L_k^\alpha(x)$ is the Sonine polynomial defined by
\begin{equation}
\sum_{k=0}^\infty z^k L_k^\alpha(x)=\frac{1
}{(1-z)^{1+\alpha}}e^{\frac{zx}{z-1}}.
 \label{eq:expansionS}
\end{equation}
It is related to the Laguerre polynomial $L_k(x)=e^x\frac{d^k}{dx^k}(e^{-x}x^k)$ via
$L_k^\alpha(x)=\frac{1}{(n+\alpha)!}%
\left(-\frac{d}{dx}\right)^\alpha L_{k+\alpha}(x)$.

\vspace*{1ex}\noindent
{\it Strong field limit. --}
In the limit of strong field $\sqrt{|QB|}\gg\max(\mu,T,m)$, only the contribution
from the lowest Landau level may be retained:
$$
\textstyle
i\bar{S}(i\omega_n,p_\parallel,\bm{p}_\perp)\to 2e^{-\ell^2\bm{p}_\perp^2}%
\frac{\slashed{p}_\parallel+\slashed{\mu}}{(i\omega_n+\mu)^2-p_\parallel^2}.
$$
This prescription is called the lowest Landau level (LLL) approximation.

\vspace*{1ex}\noindent
{\it Weak field limit. --}
On the other hand,
In the case that the magnetic field is not so strong,
we need to sum up over all Landau levels.
This can be done by introducing the proper time $s$ as follows.
Let $\mu$ be positive without loss of any generality, we have
$$
\begin{array}{rcl}
\textstyle\frac{1}{(i\omega_n+\mu)^2-2k/\ell^2-p_\parallel^2-m^2}%
  &=&\displaystyle-i\theta(\omega_n)\int_0^\infty ds %
  e^{-s(A+i2k/\ell^2)}\\[2ex]
  &&\displaystyle+i\theta(-\omega_n)\int_0^\infty ds %
  e^{+s(A+i2k/\ell^2)},
\end{array}
$$
where we have introduced $A$ just for notational simplicity as
$$
A=2\mu\omega_n+i(\omega_n^2-\mu^2+p_\parallel^2+m^2).
$$
Using this expression, the summation over the Landau levels can be
performed with the help of Eq.~(\ref{eq:expansionS}), resulting
in the expression
\begin{equation}
 \begin{array}{rcl}
 \bar{S}(p)&=&-i\int_0^\infty
  ds e^{-s A}e^{-i\ell^2\bm{p}_\perp^2\tan(s/\ell^2)}\bigg\{%
  (\slashed{p}+\slashed{\mu}+m)\\[2ex]
  &&\qquad-\mathrm{sgn}(QB)\gamma^1\gamma^2(\slashed{p}_\parallel+\slashed{\mu}+m)\tan(s/\ell^2)\\[2ex]
  &&\qquad-\bm{p}_\perp\cdot\bm{\gamma}_\perp\tan^2(s/\ell^2)\bigg\}.
   \end{array}
\end{equation}
In the weak field case where $|QB|\ll\min(\mu,T,m)$, we can expand the
integrand of above expression with respect to $1/\ell^2=|QB|$, and perform
the $s$ integration exactly.
Up to the quartic order in $\ell$ (corresponding to the quadratic order
in $|QB|$)
we have
\begin{equation}
 \begin{array}{rcl}
  \bar{S}(p)&=&S^{(0)}+(QB)S^{(1)}+(QB)^2S^{(2)}+\cdots,
\end{array}
\end{equation}
where
\begin{equation}
\begin{array}{rcl}
  S^{(0)}&=&\displaystyle%
   \frac{\slashed{p}+\slashed{\mu}+m}{(i\omega_n+\mu)^2-\bm{p}^2-m^2},\\[2ex]
   S^{(1)}&=&\displaystyle\frac{\slashed{p}_\parallel+\slashed{\mu}+m}%
  {((i\omega_n)^2-\bm{p}^2-m^2)^2}(i\gamma^1\gamma^2),\\[2ex]
  S^{(2)}&=&%
  \displaystyle\frac{2\bm{p}_\perp\cdot\bm{\gamma}_\perp}%
  {((i\omega_n+\mu)^2-\bm{p}^2-m^2)^3}%
  -\frac{2\bm{p}_\perp^2(\slashed{p}+\slashed{\mu}+m)}%
  {((i\omega_n+\mu)^2-\bm{p}^2-m)^4}.\\[2ex]
 \end{array}
 \label{eq:full}
\end{equation}
If we ignore the last term, staying at the lowest nontrivial order in the expansion,
the expression above reduces to Eq.~(\ref{eq:propinB}).

\vspace*{1ex}
\section{Weak field expansion of Ginzburg-Landau coefficients}%
\label{eq:exdemo}
Using the fermion propagator Eq.~(\ref{eq:full}), we can derive the weak
field expansion
of the GL energy functional at any order in principle, when required.
Apart from the new term Eq.~(\ref{eq:anom}) obtained at the lowest
nontrivial order in $B$, there should also be the corrections to the
original GL couplings $\alpha_2$, $\alpha_4$ and etc.
But the corrections should be at least quadratic order in $\bm{B}$
because of the rotational symmetry of the system. 
There is no vector except for $\bm{B}$ itself
which can be used to make the scalar product.
We here show this by explicit computation of the magnetic field
corrections to $\alpha_2$ and $\alpha_4$.
\subsection{Evaluation of $\alpha_2$}
The expression for $\alpha_2$, except for the model dependent constant
term, is given by
$$
\alpha_2=N_cT\sum_{f=u,d}\int_0^{1/T} d\tau d\tau_2\int d{\bf x}_2%
{\mathrm{tr}}\left[S_f(x,x_2)S_f(x_2,x)\right],
$$
where $S_f$ is the quark propagator for flavor $f=(u,d)$ in the presence of magnetic field,
$x$ and $x_2$ is understood as the four-vectors, \ie, $x=(-i\tau,{\bf x})$, $x_2=(-i\tau_2,{\bf x}_2)$.
The Schwinger phase cancels out so we can evaluate the functional trace
with the Fourier decomposition of $\bar{S}_f(x-x_2)$ as,
$$
\alpha_2=N_cT\sum_{i=u,d}\sum_{n}\int\frac{d{\bf p}}{(2\pi)^3}%
{\mathrm{tr}}\left[\bar{S}_f(p)\bar{S}_f(p)\right].
$$
If we write the expansion of $\alpha_2$ in $B$ as
$$
\alpha_2=\alpha_2^{(0)}+(eB)\alpha_2^{(1)}+(eB)^2\alpha_2^{(2)}+\cdots,
$$
using Eq.~(\ref{eq:full}), we have
$$
\begin{array}{rcl}
 \alpha_2^{(1)}&=&\displaystyle 2N_c\sum_{f=u,d}\frac{Q_f}{e}%
  T\sum_n\int\frac{d{\bf p}}{(2\pi)^3}\mathrm{tr}%
 \left[\bar{S}^{(1)}\bar{S}^{(0)}\right],\\[2ex]
 \alpha_2^{(2)}&=&\displaystyle N_c\sum_{i=u,d}\left(\frac{Q_f}{e}%
 \right)^2T\sum_n\int\frac{d{\bf p}}{(2\pi)^3}\mathrm{tr}%
 \left[\bar{S}^{(1)}\bar{S}^{(1)}+2\bar{S}^{(0)}\bar{S}^{(2)}\right],
 \end{array}
 $$
 where $Q_u=2e/3$, and $Q_d=-e/3$.
 After computing the trace over Dirac indices we see $\alpha_2^{(1)}$ is
 actually vanishing as guaranteed by the rotational symmetry.
On the other hand, we obtain
 $\alpha_2^{(2)}=\frac{5}{27}\alpha_6^{(0)}$ with the help of the
 prescription introduced \cite{Nickel:2009ke}.
As a consequence, the final expression for $\alpha_2$ up to the
 quadratic order in $B$ can be summarized as
 \begin{equation}
  \alpha_2(\mu,T)=\alpha_2^{(0)}(\mu,T)+\frac{5}{27}\big(e\bm{B}\big)^2%
   \alpha_6^{(0)}(\mu,T).
 \end{equation}
 $\{\alpha_{2n}^{(0)}\}$ are the GL coefficients in the absence of
 magnetic field, Eq.~(\ref{eq:alphas}).

Although it is beyond the scope of the present paper, let us briefly
discuss the effect of $B^2$ to the chiral restoration in the chiral limit.
It can be shown that $\alpha_6(\mu,T)$ changes its sign at some critical
value of $\mu/T$ which we denote by $c$. 
For the NJL type model we obtain $c\sim 0.5$.
For $\mu/T<c$, $\alpha_6<0$ and for $\mu/T>c$, $\alpha_6>0$.
This suggest that for low density $\alpha_2$ has a negative feedback
from the magnetic field so that the magnetic field enhances the chiral
symmetry breaking, and thus increases the critical temperature.
That is magnetic catalysis of chiral symmetry breaking.
On the other hand, for high density the $B^2$ effect increases
$\alpha_2$, so the chiral symmetry tends to be restored. The magnetic
field is expected to decrease the critical temperature.
This is the inverse magnetic catalysis.

\subsection{Evaluation of $\alpha_4$}
Similarly we expand $\alpha_4$ in powers of $eB$ as
$$
\alpha_4=\alpha_4^{(0)}+(eB)\alpha_4^{(1)}+(eB)^2\alpha_4^{(2)}+\cdots.
$$
The microscopic expression for $\alpha_4$ is
\begin{equation}
\begin{array}{rcl}
\alpha_4&=&\displaystyle N_cT\sum_{f=u,d}\int_0^{1/T}d\tau d\tau_2 
d\tau_3 d\tau_4\int d{\bf x}_2 d{\bf x}_3 d{\bf x}_4\Big\{\\[2ex]
&&\displaystyle\mathrm{tr}\left[%
S_f(x,x_2)S_f(x_2,x_3)S_f(x_3,x_4)S_f(x_4,x)\right]\Big\}.
\end{array}
\label{eq:alpha4}
\end{equation}
The effect of the Schwinger phase should be carefully examined
in this case, because it no longer vanishes. In fact
\begin{equation}
 \textstyle\frac{\Phi_f(x,x_2)+\Phi_f(x_2,x_3)+\Phi_f(x_3,x_4)+\Phi_f(x_4,x)}{Q_f}%
  =\displaystyle\oint_{\partial\bm{S}}d\bm{z}\cdot\bm{A}(\bm{z}),
\label{eq:1}
\end{equation}
where $\partial\bm{S}$ is the closed boundary which follows the 
straight paths ${\bf x}\to{\bf x}_4\to{\bf x}_3\to{\bf x}_2\to{\bf x}$.
Making use of the Stokes' theorem, we have
\begin{equation}
\oint_{\partial\bm{S}}d\bm{z}\cdot\bm{A}(\bm{z})=\bm{B}\cdot\bm{S},
\label{eq:2}
\end{equation}
where $\bm{S}$ is 
the sum of planar area vectors which all together has the boundary 
$\partial\bm{S}$. The simplest choice is the sum of two triangles,
one made by ${\bf x}_4-{\bf x}$ and ${\bf x}-{\bf x}_3$ and the other
made by ${\bf x}_3-{\bf x}$ and ${\bf x}-{\bf x}_2$, namely
\begin{equation}
\bm{S}=\frac{({\bf x}_3-{\bf x})\times({\bf x}_4-{\bf x}_2)}{2}.
\label{eq:3}
\end{equation}
Although an individual Schwinger phase is neither gauge invariant nor
translationally invariant, the above combination is properly in the invariant
form; as for translational invariance, it is seen by making the spatial
translation ${\bf x}_i\to{\bf c}$ with ${\bf c}$ being an arbitrary shift.
As a result, $\alpha_4$ does not depend on ${\bf x}$ as it should.

Plugging Eq.~(\ref{eq:1}) together with Eqs.~(\ref{eq:2}) and
(\ref{eq:3}) in to Eq.~(\ref{eq:alpha4}), and performing some
momentum integrations, we obtain
\begin{equation}
\begin{array}{rcl}
\alpha_4&=&\displaystyle 
N_cT\sum_{f=u,d}\sum_n\int\frac{d{\bf
 p}_{1}}{(2\pi)^3}\frac{d{\bf p}_{3}}{(2\pi)^3}\int d{\bf x}_3%
e^{-i({\bf p}_1-{\bf p}_3)\cdot{\bf
x}_3}\\[2ex]
&&\quad\times\mathrm{tr}\bigg[\bar{S}_f(i\omega_n,{\bf p}_1)%
\bar{S}_f\left(i\omega_n,{\bf p}_1-\frac{Q_f}{2}{\bf x}_3\times\bm{B}\right)%
\\[2ex]
&&\qquad\quad\times\bar{S}_f(i\omega_n,\bm{p}_3)\bar{S}_f%
\left(i\omega_n,{\bf p}_3+\frac{Q_f}{2}{\bf x}_3\times\bm{B}\right)\bigg].
\end{array}
\end{equation}
The momentum shifts in two propagators are originated in the Schwinger phase.
Remaining task is to evaluate the integral by expanding the integrand in powers
of $(eB)$.
For example, the propagator having the finite magnetic momentum shift
can be expanded up to the first order in $(eB)$ as follows
$$
\begin{array}{rcl}
 \bar{S}_f\left(i\omega_n,{\bf p}_1-\frac{Q_f}{2}{\bf x}_3\times\bm{B}\right)%
 &=&\bar{S}^{(0)}(i\omega_n,{\bf p}_1)+(Q_fB)\bar{S}^{(1)}(i\omega_n,{\bf p}_1)\\[2ex]
 &&-\frac{Q_f}{2}({\bf x}_3\times\bm{B})\cdot\nabla_{{\bf p}_1}\bar{S}^{(0)}(i\omega_n,{\bf p}_1).
\end{array}
$$
Collecting terms up to the order $(eB)^2$ corrections in the integrand,
and making some tedious computations, we arrive at the conclusion
that the first order correction $\alpha_4^{(1)}$ is in fact vanishing
(as expected from the rotational symmetry) and the second order correction
is
$$
\begin{array}{rcl}
 \alpha_4^{(2)}&=&2N_c\sum_f\left(\frac{Q_f}{e}\right)^2T\sum_n%
  \int\frac{d{\bf p}}{(2\pi)^3}\mathrm{tr}\bigg[2\big(\bar{S}^{(0)}\big)^3S^{(2)}%
  \\[2ex]
  &&\quad+\big(\bar{S}^{(0)}\bar{S}^{(1)}\big)^2+2\big(\bar{S}^{(0)}\big)^2\big(\bar{S}^{(1)}\big)^2\bigg]\\[2ex]
  &&-2iN_c\epsilon_{3ij}\sum_f\left(\frac{Q_f}{e}\right)^2T\sum_n\int\frac{d{\bf p}}{(2\pi)^3}%
  \mathrm{tr}\bigg[%
  \bar{S}^{(0)}\bar{S}^{(1)}\bar{S}^{(0)}\\[2ex]
  &&\quad\times\Big(\gamma^i\big(\bar{S}^{(0)}\big)^2\gamma^j\bar{S}^{(0)}%
  +\bar{S}^{(0)}\gamma^i\big(\bar{S}^{(0)}\big)^2\gamma^j%
  +\gamma^i\big(\bar{S}^{(0)}\big)^3\gamma^j\Big)\bigg]\\[2ex]
  &&-\frac{N_c}{4}\epsilon_{3ij}\epsilon_{3\ell m}\sum_f\left(\frac{Q_f}{e}\right)^2T\sum_n%
  \int\frac{d{\bf p}}{(2\pi)^3}\mathrm{tr}\bigg[
  \gamma_i\big(\bar{S}^{(0)}\big)^2\gamma_j\big(\bar{S}^{(0)}\big)^2\\[2ex]
  &&\quad\times\gamma_\ell\big(\bar{S}^{(0)}\big)^2\gamma_m\big(\bar{S}^{(0)}\big)^2%
  +\big(\bar{S}^{(0)}\big)^2\Big\{\gamma_i\bar{S}^{(0)}\gamma_\ell-\gamma_\ell\bar{S}^{(0)}\gamma_i\Big\}\\[2ex]
  &&\quad\times\big(\bar{S}^{(0)}\big)^2\Big\{\gamma_j\bar{S}^{(0)}\gamma_m-\gamma_m\bar{S}^{(0)}\gamma_j%
  \Big\}\big(\bar{S}^{(0)}\big)^2\bigg]\equiv\frac{5}{9}\alpha_8^{(0)}.
\end{array}
$$
Our results can be summarized in the following compact form:
\begin{equation}
\alpha_4(\mu,T)=\alpha_4^{(0)}(\mu,T)+\frac{5}{9}\big(e\bm{B}\big)^2\alpha_8^{(0)}(\mu,T).
\end{equation}

\end{document}